\RequirePackage{fix-cm}
\documentclass[smallextended]{svjour3}       
\usepackage[italian,english]{babel}
\smartqed  
\usepackage{amsmath}
\usepackage{amssymb}
\usepackage{lipsum}
\usepackage{dsfont}
\usepackage{multirow}
\usepackage{afterpage}
\usepackage{amsfonts}
\usepackage{bm}%
\usepackage{babel}
\usepackage{hyperref}
\usepackage{graphicx}
\usepackage{color}
\usepackage{graphicx}
\begin{document}

\title{Spotlighting quantum phase transition in spin -1/2 Ising–Heisenberg diamond chain employing
Measurement-Induced Nonlocality 
}


\author{S. Bhuvaneswari \and R. Muthuganesan       \and    R. Radha 
}


\institute{Centre for Nonlinear Science (CeNSc),  Government College for Women, Kumbakonam - 612 001, Tamil Nadu, India \at
              \email{bhuvanajkm85@gmail.com}           
            \and
         Centre for Nonlinear Science \& Engineering, School of Electrical \& Electronics Engineering,
SASTRA Deemed University, Thanjavur, Tamil Nadu 613 401, India\at
              \email{rajendramuthu@gmail.com}
              \and
           Centre for Nonlinear Science (CeNSc),  Government College for Women, Kumbakonam - 612 001, Tamil Nadu, India \at
           \email{vittal.cnls@gmail.com}
}

\date{Received: date / Accepted: date}

\maketitle

\begin{abstract}
We examine thermal quantum correlations characterized by Measurement-Induced Nonlocality (MIN) in an infinite spin-1/2 Ising-Heisenberg spin chain with Dzyaloshinskii-Moriya (DM) interaction. We evaluate MIN analytically in the thermodynamic limit using the transfer matrix approach and show that the MIN and its first-order derivative may spotlight the quantum criticality and quantum phase transition (QPT). We observe that the DM interaction reduces the role of anisotropy parameter in  initiating QPT. Further, the DM interaction also induces the nonlocality in the system if the spins are unentangled and greatly enhances the quantum correlations if the spins are correlated.  The impact of the magnetic field and temperature on quantum correlations is also brought out at a critical point.

\end{abstract}

\section{Introduction}
Nonlocality, a unique and fundamental characteristic  feature of composite quantum systems \cite{Einstein,Schrodinger1,Schrodinger2}, has been proven to be a useful  resource for real-life applications, notably in the domain of  secure communication, cryptography, etc. In the realm of Bell nonlocality  \cite{Bell}, the violation of any Bell inequality can manifest itself as nonlocality   and entanglement can be recognized as  one of the most important  signatures of nonlocal aspects of quantum states. Any pure entangled two-qubit state violates the Clauser, Horne, Shimony, and Holt (CHSH) inequality, while it is not true for mixed states \cite{Almeida}. Werner showed that the entanglement is not a complete picture of nonlocality \cite{Werner}. Further, the deterministic quantum computation with one qubit (DQC1) \cite{Datta2008} protocol is demonstrated through the separable state (zero entanglement), implying that there exists a special kind of property beyond entanglement, which is also efficient  in solving some classically intractable problems. In recent times, the detailed investigation on these domains reveal that the entanglement does not measure the quantum correlations present in the quantum system completely \cite{Datta2008}. To address this issue, various measures  have been identified to measure the quantum  correlations which can not be grasped by the entanglement \cite{Ollivier2001,Luo2008PRA,Dakic2010,Luo2011,UIN2014}.
 
In particular, Luo and Fu introduced a new measure called Measurement-Induced Nonlocality (MIN) \cite{Luo2011},   which is based on the fact that local disturbance due to von Neumann projective measurements on marginal state can influence globally. Under these local measurements, the marginal states are invariant, and due to the local invariance, MIN is considered to  be a secure resource of quantum communications and cryptography. Originally, this quantity is defined in terms of maximal Hilbert-Schmidt distance between the pre- and post-measurement states and is considered to be more general than that of Bell’s version of nonlocality. It is well-known that the correlation  quantifiers based on the Hilbert-Schmidt distance are not a faithful measure \cite{Piani2012}.  In order to resolve the local ancilla problem, different forms of MIN  have been introduced \cite{HFan2015,MuthuPLA,Muthu2,XiMIN,HuMIN,HFanReview}
 
Quantum phase transition (QPT) is defined as a transition between distinct ground states of quantum many-body systems when a controlled  or tuning parameter in the Hamiltonian crosses a critical point. In general, QPT can be observed at low  temperatures when quantum fluctuations dominate the thermal effects \cite{Sachdev}. The condensed matter systems  can be considered as a natural playground  for manipulating quantum information.  The natural mineral azurite $Cu_3(CO_3)_2(OH)_2$ is an interesting quantum antiferromagnetic system described by the Heisenberg model on a generalized diamond chain \cite{Honecker2011}. Due to its experimental realization and spectacular magnetic properties, the  diamond chain  has  caught the attention of physicists working on quantum information theory. In recent times, a lot of attention has been focused on understanding the connection between  quantum information theory and condensed matter systems. Entanglement not only captures quantum correlations, but also detects phase transition in physical systems \cite{Osterloh2002,Osborne2002,Kargarian2007,Kargarian2007a,Ma2011a,Ma2011b,Xu2013,Preskill
}. It is found that quantum discord is also helpful  in the identification of phase transition \cite{Dillenschneider,Sarandy2009,Maziero2012,Yao2012,Cheng2012,WerlangPRA2010,WerlangPRL2010,WerlangPRA2011,
Werlang2013,Li2011,Maziero2010,Cakmak2012}. 

In the present paper, we consider the spin-1/2 Ising-XXZ diamond  chain and construct  transfer matrix. We then identify the trace distance based measurement-induced nonlocality as an indicator of QPT in Ising-XXZ diamond chain. We evaluate MIN analytically and demonstrate how this measure could be useful in detecting QPT in Ising-XXZ diamond spin chain. Further, the role of DM interaction and magnetic field on quantum correlations and QPT is also highlighted.

\section{Measurement-induced nonlocality}

 In recent times, Measurement-Induced Nonlocality  (MIN) has been employed as quantum correlation quantifiers. MIN is defined as the maximal nonlocal and global  effect due to locally invariant, originally captured via the maximal Hilbert-Schmidt distance between the quantum state of a bipartite system and the corresponding state after performing a local measurement on one subsystem which does not change the state of this subsystem. Mathematically, it is defined as \cite{Luo2011}
\begin{equation}
 N_2(\rho) =~^{\text{max}}_{\Pi ^{a}}\| \rho - \Pi ^{a}(\rho )\|^{2}_2 \label{MIN2}
\end{equation}
where the maximization is taken over the von Neumann projective measurements on subsystem $a$ and  $\Vert \mathcal{O} \Vert_2 = \sqrt{\text{Tr}\mathcal{O}^{\dagger}\mathcal{O}}$ is the Hilbert-Schmidt norm. Here $\Pi^{a}(\rho) = \sum _{k} (\Pi ^{a}_{k} \otimes   \mathds{1} ^{b}) \rho (\Pi ^{a}_{k} \otimes  \mathds{1}^{b} )$, with $\Pi ^{a}= \{\Pi ^{a}_{k}\}= \{|k\rangle \langle k|\}$  being the projective measurements on the subsystem $a$, which  does not change the marginal state $\rho^{a}$ locally i.e., $\Pi ^{a}(\rho^{a})=\rho ^{a}$. If $\rho^{a}$ is non-degenerate, then the maximization is not required.  This quantity is easy to compute  and realizable. However, Piani pointed out that the Hilbert-Schmidt based quantity is not a valid measure of quantum correlation \cite{Piani2012}.

To resolve this issue, a reliable geometric quantifier of correlation measure is defined through the Schatten 1-norm distance (trace norm) \cite{HFan2015}, which naturally fixes the local ancila problem \cite{Piani2012}. It is defined as
\begin{equation}
N_1(\rho):= ~^{\text{max}}_{\Pi^a}\Vert\rho-\Pi^a(\rho)\Vert_1
\end{equation} 
where $\Vert \mathcal{O} \Vert_1 = \text{Tr}\sqrt{\mathcal{O}^{\dagger}\mathcal{O}}$ is the trace norm of operator $\mathcal{O}$. Here also, the maximization is taken over all von Neumann projective measurements.


\section{The model}\label{Sec3}
\begin{figure*}[!ht]
\centering\includegraphics[width=0.6\linewidth]{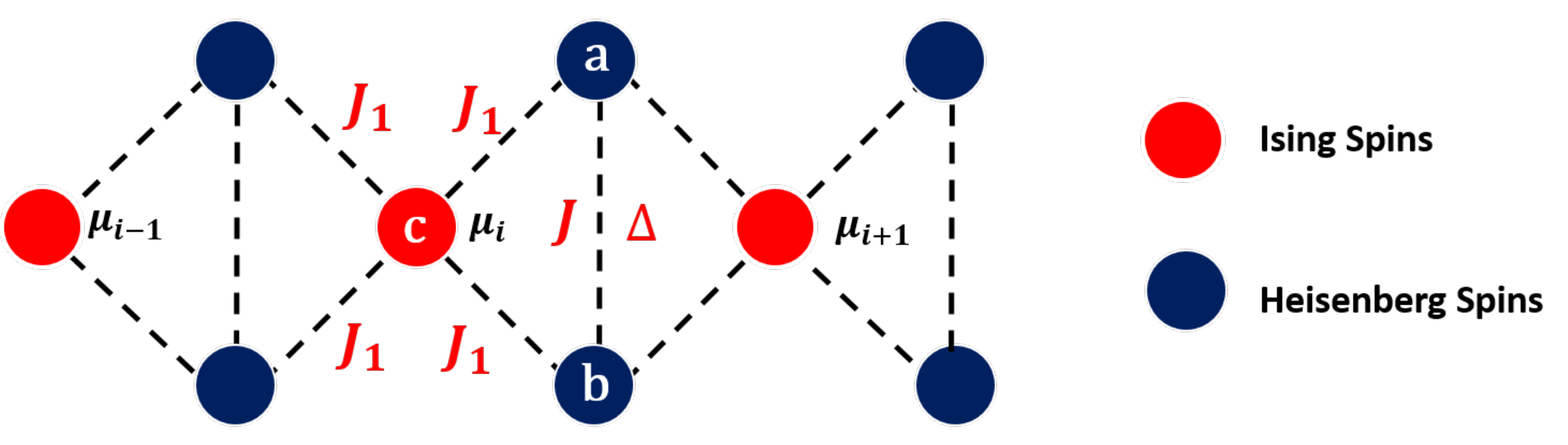}
\caption{(color online) Schematic representation of Ising-XXZ diamond spin chain}
\label{fig}
\end{figure*}

In this section, we introduce the Hamiltonian of the spin-1/2 Ising-XXZ model on a diamond chain with Dzyaloshinskii-Moriya (DM) interaction $D$ in the presence of an external magnetic field $h$. The model consists of the interstitial  Heisenberg spins $\left({\bf S}^a_i, {\bf S}^b_{i}\right)_\Delta$ and Ising spins located in the nodal site. Fig. (\ref{fig}) exemplifies the schematic representation of Ising-XXZ diamond spin chain. The total Hamiltonian of the model can be written as
 \begin{align}
 \mathcal{H}=\sum_i^N\mathcal{H}_i
 \end{align}
 with  $i^{\text{th}}$ block Hamiltonian  being
 \begin{align}
 \mathcal{H}_i=&J\left({\bf S}^a_i, {\bf S}^b_{i}\right)_\Delta+J_1\left(S^a_{z,i}+S^b_{z,i}\right)(\mu_i+ \mu_{i+1})+{\bf D} \cdot \left({\bf S}^a_i\times {\bf S}^b_i \right)\nonumber \\
 &-h\left(S^a_{z,i}+ S^b_{z,i}\right)  -\frac{h}{2}(\mu_i+ \mu_{i+1}) \nonumber
 \end{align}
 where $\left({\bf S}^a_i, {\bf S}^b_{i}\right)_\Delta=S^a_{x,i}S^b_{x,i}+S^a_{y,i}S^b_{y,i}+\Delta S^a_{z,i}S^b_{z,i}$ with  $S^k_{\alpha,i}=\frac{\hbar}{2}\sigma_{\alpha,i} ~~ (k=a,b; ~\alpha=x,y,z)$ and $\mu_{z,i}$ are the respective components of the spin -1/2 operator. Here  $J$ stands for interaction strength between the nearest neighbour Heisenberg spins, the parameter $J_1$ corresponds to interaction between nearest Ising and Heisenberg spins, $\Delta$ is the anisotropy parameter along z axis and ${\bf D}=(0~0~D)$ is DM vector  along $z$ direction. For simplicity, we consider periodic boundary condition i.e., $\mu_{N+1}=\mu_1$. 
 In order to evaluate the spectrum of the Hamiltonian, we fix the values of $\mu_i$ and $\mu_{i+1}$. In the standard two qubit computational basis $\{ \lvert 00\rangle,\lvert 01\rangle,\lvert 10\rangle,\lvert 11\rangle\}  $,  the eigenvalues and their corresponding eigenvectors of the block Hamiltonian $\mathcal{H}_i$ are computed as 
 \begin{align}
\lambda_1 \left(\mu_i, \mu_{i+1} \right)&=\frac{1}{4}J\Delta- \left(\frac{h}{2}-J_1\right) \left(\mu_i+ \mu_{i+1} \right)-h,~~~~~|\varphi_1\rangle=  |00\rangle, \nonumber \\
\lambda_2 \left(\mu_i, \mu_{i+1} \right)&=-\frac{\eta}{2}-\frac{1}{2}h\left(\mu_i+ \mu_{i+1} \right)-\frac{1}{4}J\Delta,~~~~~~~~~~~~ |\varphi_2\rangle=  \frac{1}{\sqrt{2}}\left(-a|01\rangle+|10\rangle \right), \nonumber \\
\lambda_3 \left(\mu_i, \mu_{i+1} \right)&=\frac{\eta}{2}-\frac{1}{2}h\left(\mu_i+ \mu_{i+1} \right)-\frac{1}{4}J\Delta,~~~~~~~~~~~~~~|\varphi_3\rangle=  \frac{1}{\sqrt{2}}\left(a|01\rangle+|10\rangle \right), \nonumber \\
\lambda_4 \left(\mu_i, \mu_{i+1} \right)&=\frac{1}{4}J\Delta- \left(\frac{h}{2}+J_1\right) \left(\mu_i+ \mu_{i+1} \right)+h,~~~~~ |\varphi_4\rangle=  |11\rangle. \nonumber
\end{align}
 where $\eta=\sqrt{D^2+J^2}$ and $a=\mathrm{i}\eta/(D+\mathrm{i}J)$. 
\section{Transfer-matrix (TM) approach}\label{Sec5}
In order to study the quantum correlations and phase transition, we first obtain a partition function for a diamond chain. This model can be solved exactly using the transfer--matrix (TM) approach \cite{Rojas2012}. In order to introduce TM approach, we will define the following operator as a function of Ising spin particles $\mu_i$ and $\mu_{i+1}$:
\begin{align}
\varrho (\mu_i, \mu_{i+1})=\mathrm{e}^{-\beta \mathcal{H}_{i}(\mu_i, ~\mu_{i+1})}
\end{align}
where $\mathcal{H}_i$ corresponds to the $i^{\text{th}}$ block Hamiltonian which depends on the  neighbouring Ising spins $\mu_i$ and $\mu_{i+1}$, $\beta=1/k_B T$,  wherein $k_B$ is the Boltzmann’s constant which is
considered as unity for simplicity and $T$ is the absolute temperature. Alternatively, the spectral decomposition can be written as 
\begin{align}
\varrho (\mu_i, ~\mu_{i+1})=\sum_{i=1}^{4}\mathrm{e}^{-\beta \lambda_{i}(\mu,~\mu_{i+1})}|\varphi_i\rangle\langle\varphi_i|
\end{align}
Straightforwardly, we can obtain the Boltzmann factor by tracing out over the two--qubit operator 
\begin{align}
w(\mu_i, ~\mu_{i+1})=\text{Tr}_{ab}[ {\varrho}(\mu_i, ~\mu_{i+1})]=\sum_{i=1}^{4}\mathrm{e}^{-\beta \lambda_{i}(\mu_i, ~\mu_{i+1})}.
\end{align}

The Ising-XXZ diamond chain partition function can be written in terms of Boltzmann factor as,  
\begin{align}
Z_N=\sum_{\{\mu\}} w(\mu_1, \mu_2) \cdots w(\mu_N, \mu_1).
\end{align}
Using the transfer-matrix notation, we can write the partition function of the diamond chain straightforwardly  as $Z_N = \text{Tr}(W^N)$ with the transfer matrix being expressed as
\begin{align}
W=
\begin{bmatrix}
\omega\left(\frac{1}{2},\frac{1}{2}\right) & \omega\left(\frac{1}{2},-\frac{1}{2}\right)\\
\omega\left(-\frac{1}{2},\frac{1}{2}\right) & \omega\left(-\frac{1}{2},-\frac{1}{2}\right)
\end{bmatrix}.
\end{align}
where the transfer-matrix elements are denoted by $\omega_{++}=\omega\left(\frac{1}{2},\frac{1}{2}\right)$,  $\omega_{+-}=\omega\left(\frac{1}{2},-\frac{1}{2}\right)$ and  $\omega_{--}=\omega\left(-\frac{1}{2},-\frac{1}{2}\right)$. The eigenvalues of above transfer--matrix are 
\begin{align}
\Lambda _{\pm}=\frac{\omega _{++}+\omega _{--}\pm Q}{2}
\end{align}
where  $Q=[(\omega _{++}-\omega _{--})^2+4\omega^2_{+-}]^{1/2}$. Therefore, the partition function for finite chain under periodic boundary
conditions is given by
\begin{align}
Z_N=\Lambda_{+}^{N} +\Lambda_{-}^{N}   
\end{align}
and in the thermodynamical limit, the above function reduces to $Z_N=\Lambda_{+}^{N}$. To calculate the two-qubit Heisenberg operator bonded by Ising particles $\mu_i$ and ~$\mu_{i+1}$, we assume that the Ising spin of the particle is fixed. Thus, the qubits operator elements in the natural basis becomes
\begin{align}
\varrho=
\begin{bmatrix}
\mathcal{\varrho }_{1,1} & 0 & 0 & 0\\
0 & \mathcal{\varrho}_{2,2} & \mathcal{\varrho}_{2,3} & 0 \\
0 & \mathcal{\varrho}_{3,2} & \mathcal{\varrho}_{3,3} & 0 \\
0 & 0 & 0 & \mathcal{\varrho}_{4,4}
\end{bmatrix}
\end{align}
and the matrix elements are 
\begin{align}
\mathcal{\varrho}_{1,1}(\mu ,\mu_{i+1})=&\mathrm{e}^{-\beta \lambda_{1}(\mu ,~ \mu_{i+1})}   \notag,\\
\mathcal{\varrho}_{2,2}(\mu ,\mu_{i+1})=&\frac{1}{2}\left(\mathrm{e}^{-\beta \lambda_{2}(\mu ,~ \mu_{i+1})}+\mathrm{e}^{-\beta \lambda_{3}(\mu ,~ \mu_{i+1})}\right) \notag,\\
\mathcal{\varrho}_{2,3}(\mu ,\mu_{i+1})=&\frac{1}{2}\left(\mathrm{e}^{-\beta\lambda_{2}(\mu ,~ \mu_{i+1})}-\mathrm{e}^{-\beta \lambda_{3}(\mu ,~ \mu_{i+1})}\right) \notag,\\
\mathcal{\varrho}_{4,4}(\mu ,\mu_{i+1})=&\mathrm{e}^{-\beta \lambda_{4}(\mu ,~ \mu_{i+1})}. \nonumber
\end{align}
The thermal average for each two-qubit Heisenberg operator will be used to construct the reduced density operator. The reduced density matrix elements are given as
\begin{align}
\mathcal{\rho}_{i,j}=\frac{1}{Z_{N}}\sum_{\{\mu\}} w(\mu_1, \mu_2)\cdots w(\mu_{r-1}, \mu_r)\mathcal{\varrho}_{i,j}(\mu_r,\mu_{r+1})w(\mu_{r+1},\mu_{r+2})\cdots w(\mu_N, \mu_1).
\end{align}
Using transfer-matrix notation, the alternate definition of reduced matrix elements are 
\begin{align}
\mathcal{\rho}_{i,j}=\frac{1}{Z_{N}}Tr(W^{r-1}P_{i,j}W^{N-r})=\frac{1}{Z_{N}}Tr(P_{i,j}W^{N-1})
\end{align}
where 
\begin{align}
P_{i,j}=
\begin{bmatrix}
\mathcal{\varrho}_{i,j}\left(\frac{1}{2},\frac{1}{2}\right) & \mathcal{\varrho}_{i,j}\left(\frac{1}{2},-\frac{1}{2}\right)\\
\mathcal{\varrho}_{i,j}\left(-\frac{1}{2},\frac{1}{2}\right) & \mathcal{\varrho}_{i,j}\left(-\frac{1}{2},-\frac{1}{2}\right)
\end{bmatrix}.
\end{align}



Real systems are well represented in the thermodynamic limit $(N\rightarrow \infty )$. Hence, the reduced density operator elements after some algebraic manipulation becomes
\begin{align}
\mathcal{\rho}_{i,j}=\frac{1}{\Lambda _{+}}\bigg\{\frac{\mathcal{\varrho}_{i,j}\left(\frac{1}{2},\frac{1}{2}\right)+\mathcal{\varrho}_{i,j}\left(-\frac{1}{2},-\frac{1}{2}\right)}{2}+\frac{2\mathcal{\varrho}_{i,j}\left(\frac{1}{2},-\frac{1}{2}\right)w_{+-}}{Q}\notag\\
+\frac{\left[\mathcal{\varrho}_{i,j}\left(\frac{1}{2},\frac{1}{2}\right)-\mathcal{\varrho}_{i,j}\left(-\frac{1}{2},-\frac{1}{2}\right)\right](w_{++}-w_{--})}{2Q}\bigg\}
\end{align}
 All the elements of reduced density operator immersed on a diamond chain are
\begin{align}
\mathcal{\rho}(T)=
\begin{bmatrix}
\mathcal{\rho}_{1,1} & 0 & 0 & 0\\
0 & \mathcal{\rho}_{2,2} & \mathcal{\rho}_{2,3} & 0 \\
0 & \mathcal{\rho}_{3,2} & \mathcal{\rho}_{3,3} & 0 \\
0 & 0 & 0 & \mathcal{\rho}_{4,4}
\end{bmatrix}.
\label{Thermal}
\end{align}
It is  worth noting at this juncture that the reduced density operator is the thermal average two-qubit Heisenberg operator immersed in the diamond chain and it can be verified that $\text{Tr}\rho= 1$. For the above density matrix, the MINs  $N_1(\rho)$ and  $N_2(\rho)$ are computed as
\begin{align}
N_2(\rho)=2~ \lvert \mathcal{\rho}_{2,3}\rvert^2 ~~ \text{and} ~~ N_1(\rho)=2~ \lvert \mathcal{\rho}_{2,3}\rvert.
\end{align}
It is observed that both measures are proportional to each other and  identical  in  measuring  the   correlations. The quantities are related as $N_2(\rho)=N_1(\rho)^2/2$. Since, $N_2(\rho)$ is not a faithful measure of  quantum correlations,  we concentrate our investigation on $N_1(\rho)$ alone.

\section{Results and Discussion}\label{Sec5}
It is  quite well-known that the system  under our investigation exhibits three different magnetic phases such as a frustrated (FRU) state, a ferrimagnetic (FIM) state, and a ferromagnetic (FM) state \cite{Canova2006,Rojas2011}.  Further, in terms of entanglement, there are two phases namely, entangled and unentangled  states \cite{Rojas2012}. It is believed that entanglement is an incomplete manifestation of nonlocality. To capture complete picture of nonlocal aspects of the thermal system given in Eq. (\ref{Thermal}), we employ trace distance MIN as a correlation quantifier.
\begin{figure*}[!ht]
\centering\includegraphics[width=0.4\linewidth]{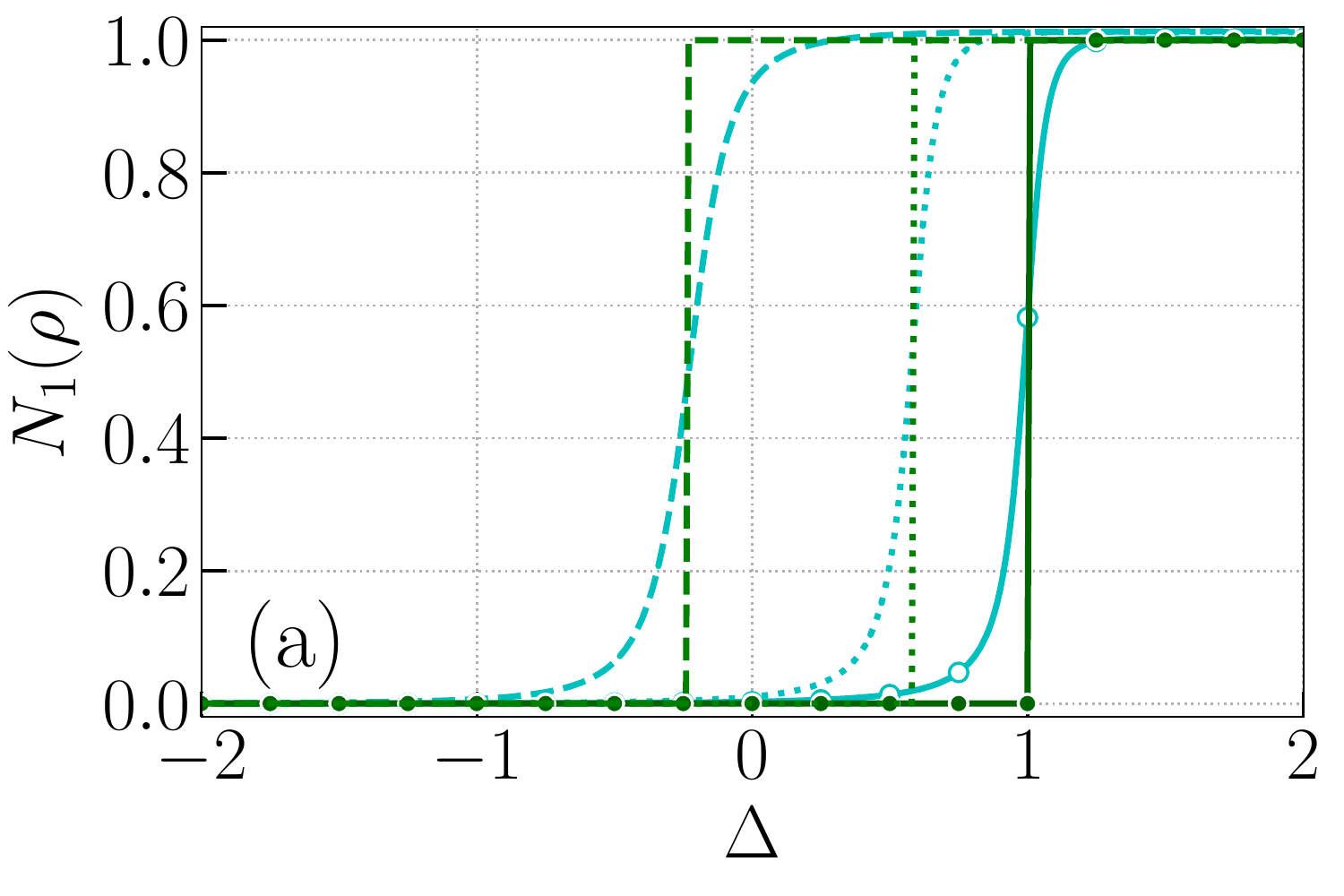}
\centering\includegraphics[width=0.4\linewidth]{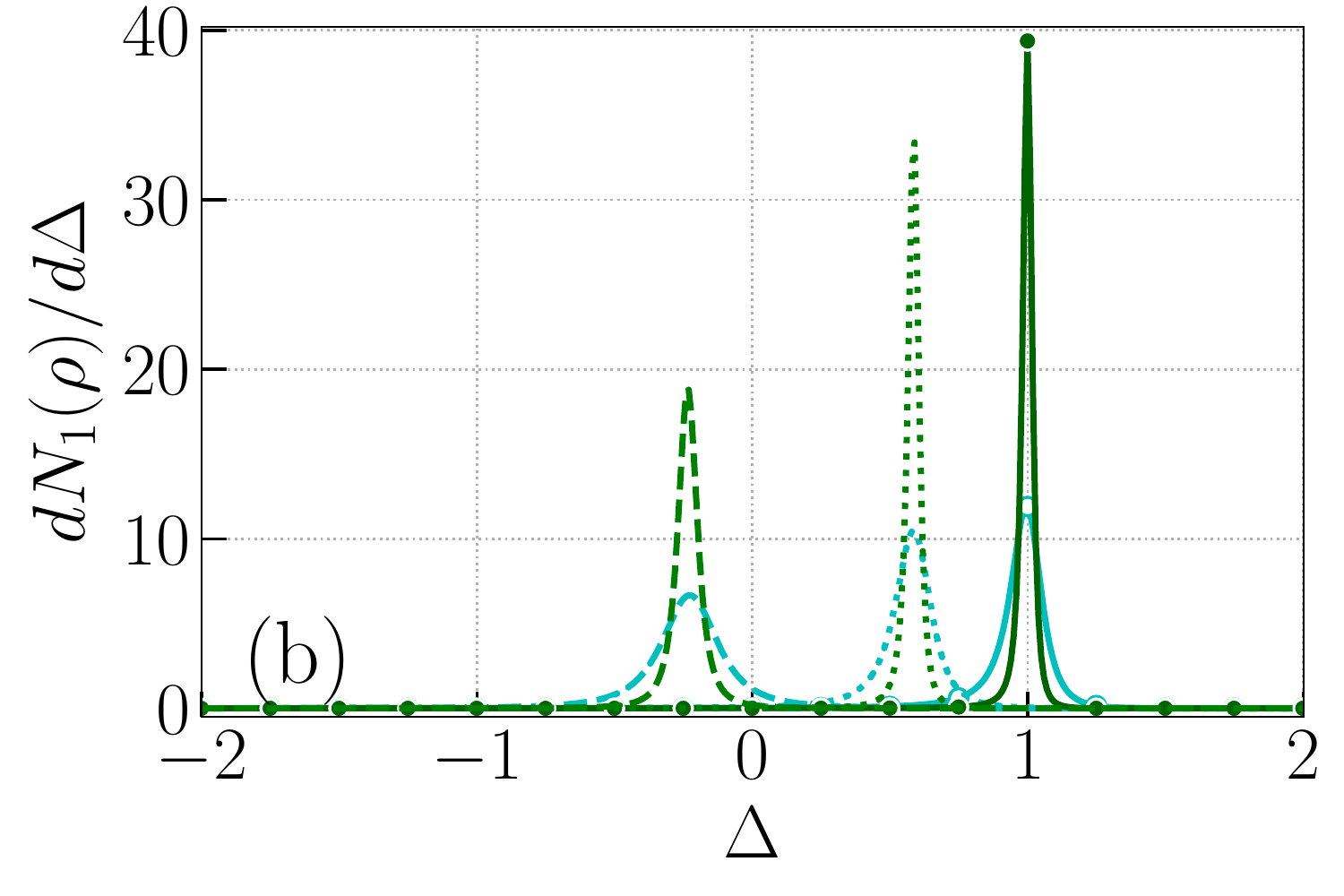}
\centering\includegraphics[width=0.4\linewidth]{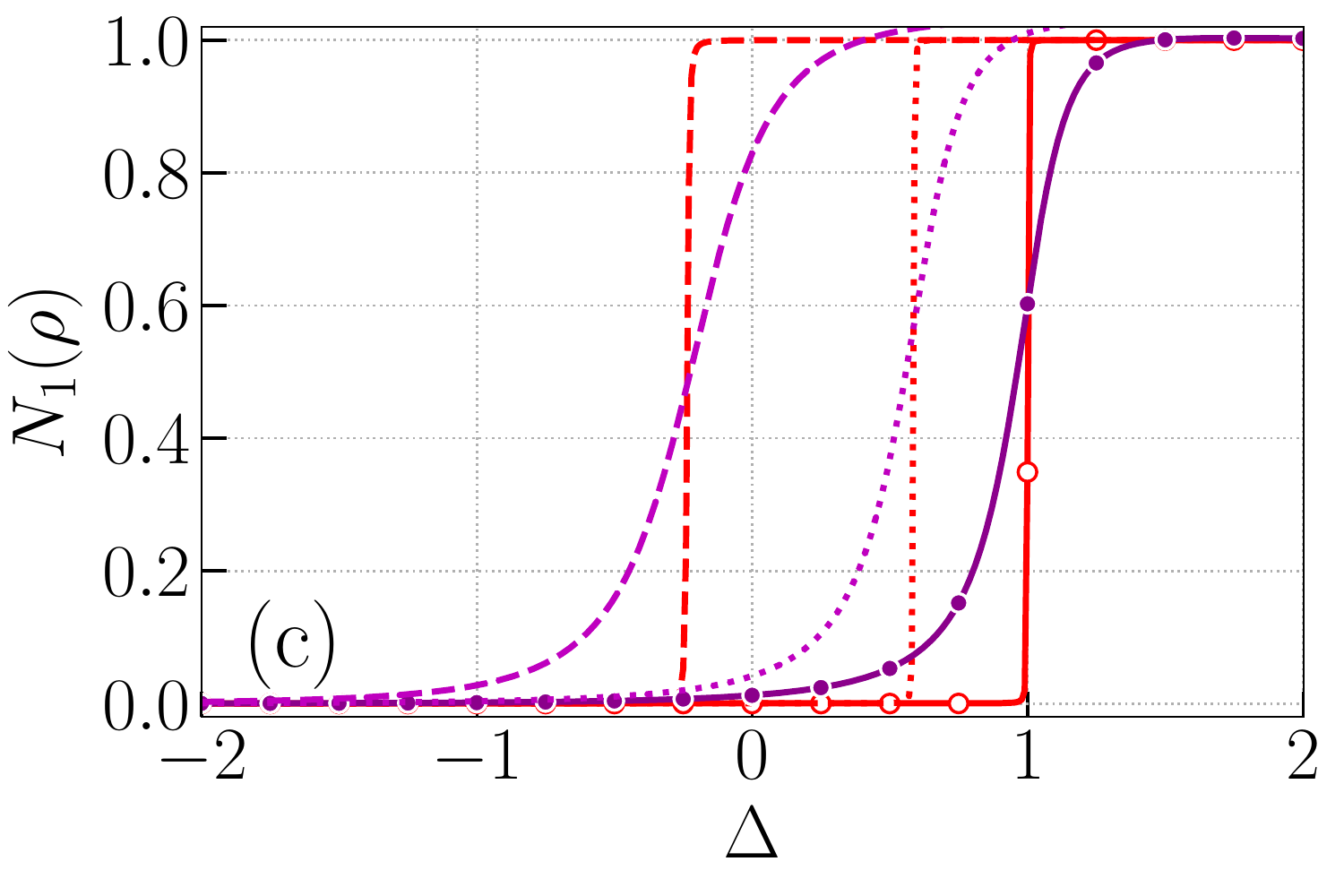}
\centering\includegraphics[width=0.4\linewidth]{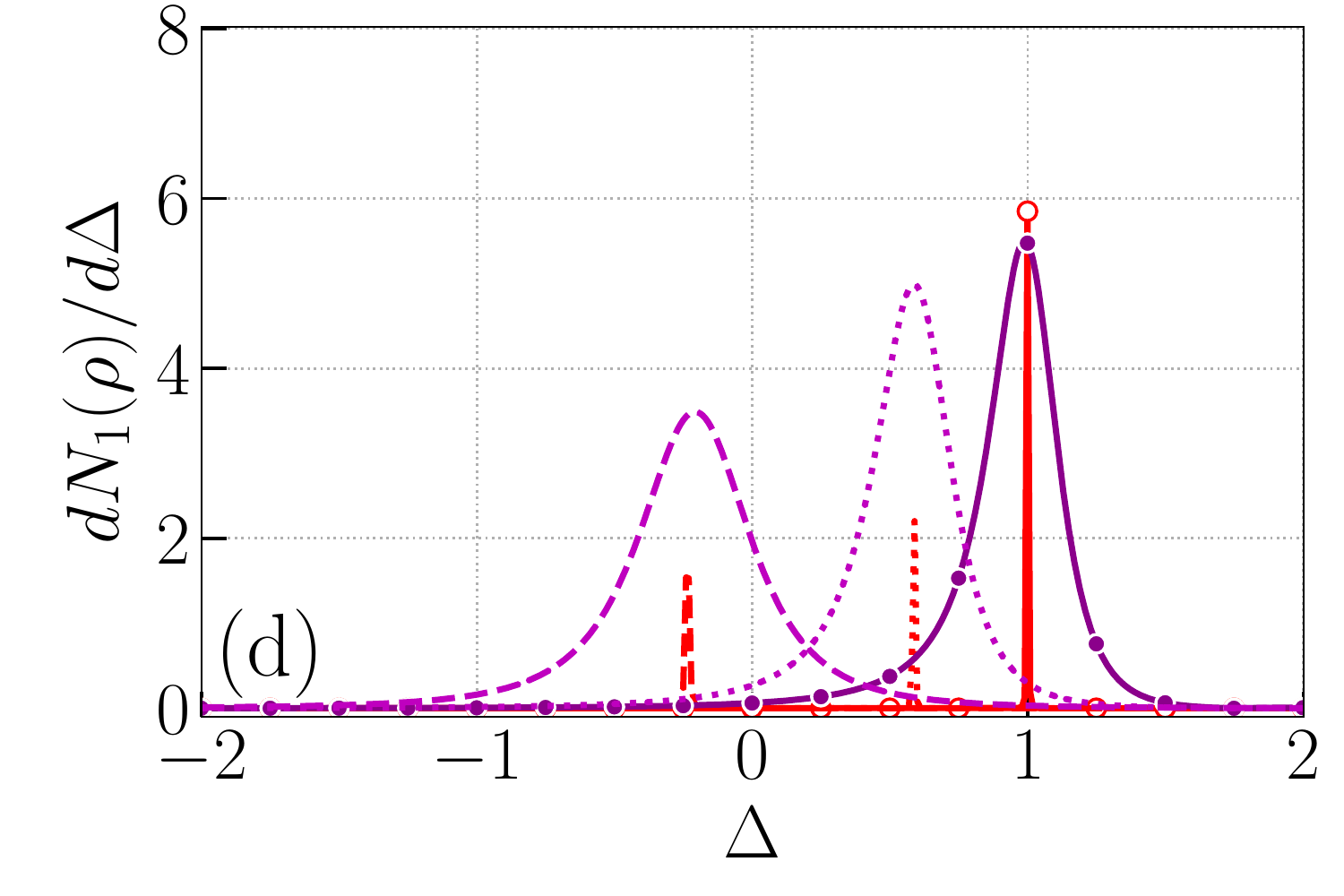}
\caption{(color online) Variation of MIN as a function of $\Delta$ with fixed parameter $J=J_1=h=1$ for different values of DM interaction  such as $D=0~ \text{(solid curve)}$, $D=1~ \text{(dotted curve)}$ and $D=2~ \text{(dashed curve)}$ (a) $T=0.15$ (green curves), $T=0.2$ (cyan curves) (c) $T=0.1$ (red curves), $T=0.25$ (purple curves) and (b) \& (d) are corresponding derivatives of MIN with respect to $\Delta$. For better visibility, we have scaled the magnitude of derivative of MIN with respect to $\Delta$.}
\label{fig1}
\end{figure*}
In order to understand the behaviors of measurement-induced correlation, we plot the trace distance based MIN in Fig. (\ref{fig1}) as a function of anisotropy parameter $\Delta$ for various values of temperatures in the absence of DM interaction and for fixed parameters $(J=J_1=h=1)$. In Rojas et al. \cite{Rojas2012}, it is shown that phase boundary between UFI (unentangled ferrimagnetic phase) and ENQ (entangled ferrimagnetic phase) is limited by $\Delta =1$,  when $h \geq 2$, whereas  the boundary between ENQ phase to UFM (unentangled ferromagnetic) phase is situated  around critical line $\Delta -3h+3=0$  with $h< 2$. For $T=0.1 ~ (\text{red curve}) $ and $0.15 ~(\text{green curve})$, it is clear that $N_1(\rho)=0$ for $\Delta \leq   1$. At  $\Delta = 1$, the MIN sharply  reaches the maximal value $1$ and  if we increase the anisotropy further  $(\Delta > 1)$, the MIN saturates at 1, which corresponds to maximally entangled state. We notice that $\Delta=1$ line separates the  unentangled state in ferrimagnetic phase (UFI) and entangled state in ferrimagnetic phase (ENQ).  Then means that $N_1(\rho)$ can signal the critical point and  phase transition at $\Delta_c=1$.  

The first derivative of MIN is also plotted in Fig. (\ref{fig1}) as a function of $\Delta$ for different values of temperatures. We  observe that the $dN_1(\rho)/d\Delta$ has the maximal value around the critical point. In other words, the derivative exhibits a clear singularity around the critical point $\Delta_c=1$ at a finite temperature. Usually, the appearance of non-analytic behavior of a physical quantity at $\Delta=1$ is  a characteristic feature of QPT.  The line $\Delta=1$ demarcates UFI and ENQ. Our observation is  similar to the phase transition predicted by the entanglement and discord \cite{Rojas2012,Gao2015}. At higher  temperatures, we also observe that increasing the temperature tends to reduce the amount of quantum correlations in the system. As temperature increases, the critical points  shift towards lower values of $\Delta$ and MIN reaches the maximal value for sufficiently larger values of $\Delta$.  Here, we observe that the transition rate between the UFI and ENQ decreases with increase of temperature due  to the smooth variation of MIN. 

Now, to analyze the effects of DM interaction on QPT and quantum correlation, we have plotted MIN and its first-order derivative as a function of $\Delta$ for different values of DM interaction ($D=1$ (dotted curve) and $D=2$ (dashed line)) in Fig.(\ref{fig1}). It is obvious that MIN values jump suddenly from zero to nonzero values earlier than $\Delta=1$ in the presence of DM interaction. Due to the introduction of DM interaction, the critical point $\Delta_c$ shifts to lower values of $\Delta$, implying that the  DM interaction reduces the role anisotropy in initiating QPT. Further, we also observe that nonlocality is  essentially induced by the DM interaction in the system.  Recently, the phase transition in the diamond chain is studied using geometric discord in the absence of DM interaction \cite{Cheng2017}. If we include the DM interaction, it is obvious that geometric discord also signals the QPT at a finite temperature for $\Delta<1$.

The above analysis with  and without DM  interaction shows that the derivative of MIN can exhibit an advantage in detecting the QPT. In other words, one can confirm the manifestation of MIN in QPT at a critical point at finite temperatures in this model.
\begin{figure*}[!ht]
\centering\includegraphics[width=0.7\linewidth]{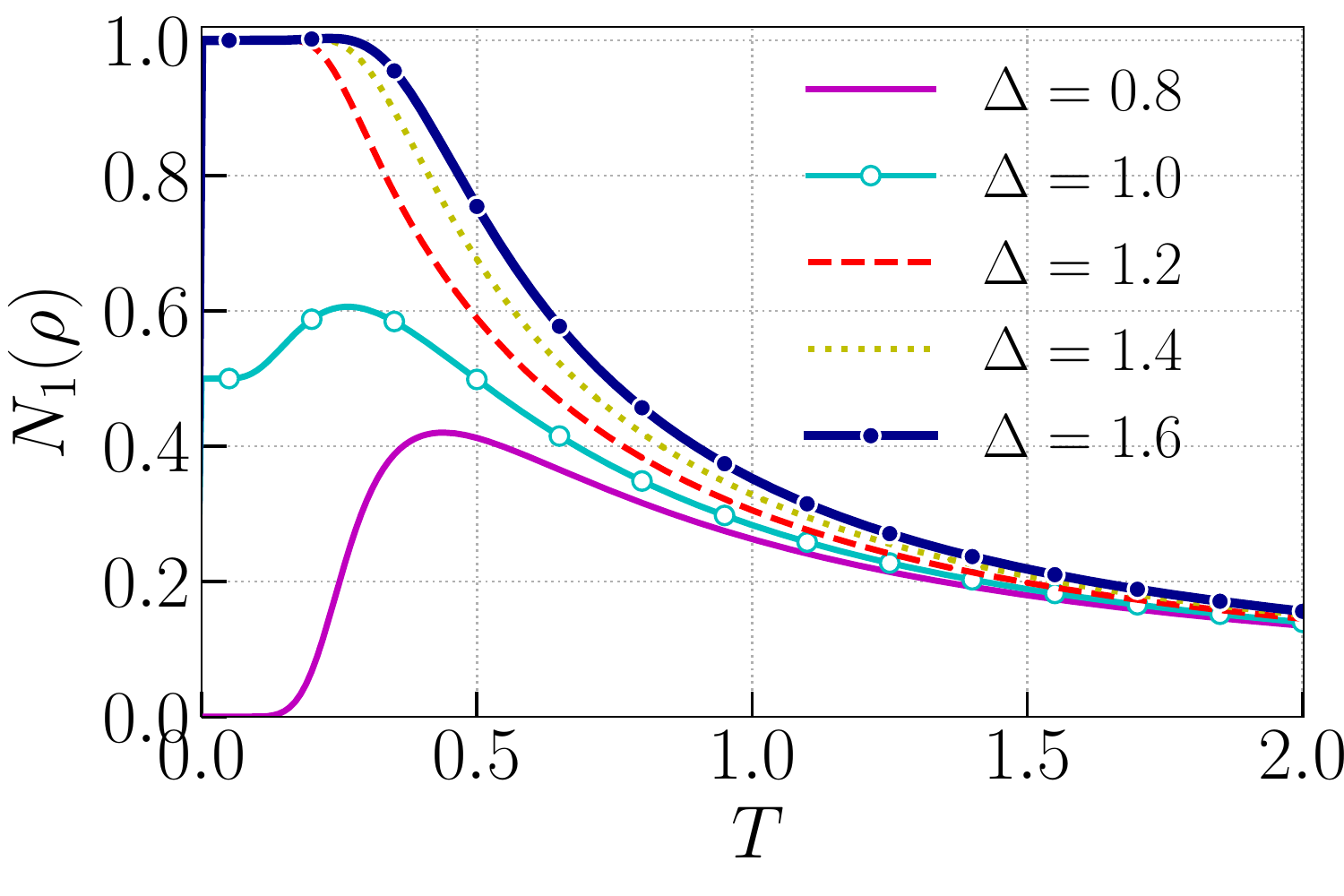}
\caption{(color online) $N_1(\rho)$ as a function of temperature at  different values of parameter $\Delta$  with fixed values of the coupling parameters $J = J_1 = 1$ and the external magnetic field $h = 1$.}
\label{fig3}
\end{figure*}

Next, we analyze the effects of temperature on quantum correlation. Here again, we show that around $\Delta=1$, the MIN can signal the critical behavior. In other words, MIN shows different behavior for $\Delta > 1$, when compared to the parametric region $\Delta \leq 1$. In Fig. (\ref{fig3}), we have plotted MIN as a  function of temperature for various values of fixed $\Delta$. As discussed above, the MIN separates  a zero correlation and a non-zero correlation state at the critical point $\Delta_c=1$. If $\Delta > 1$, the quantum correlation decreases monotonically and drops to zero in the higher temperature region. On the other hand, for $\Delta <  1$, MIN increases with temperature and reaches a maximal  value. It then decreases monotonically with temperature $T$ and disappears in the higher temperature region.  The system is in an unentangled paramagnetic phase. The entanglement tends to disappear even for low temperatures \cite{Gao2015} and MIN is more robust against thermal agitation compared to entanglement. The above results suggest that the different behaviors of MIN indicate that the system undergoes a QPT when $\Delta$ goes across 1.

\begin{figure*}[!ht]
\centering\includegraphics[width=0.4\linewidth]{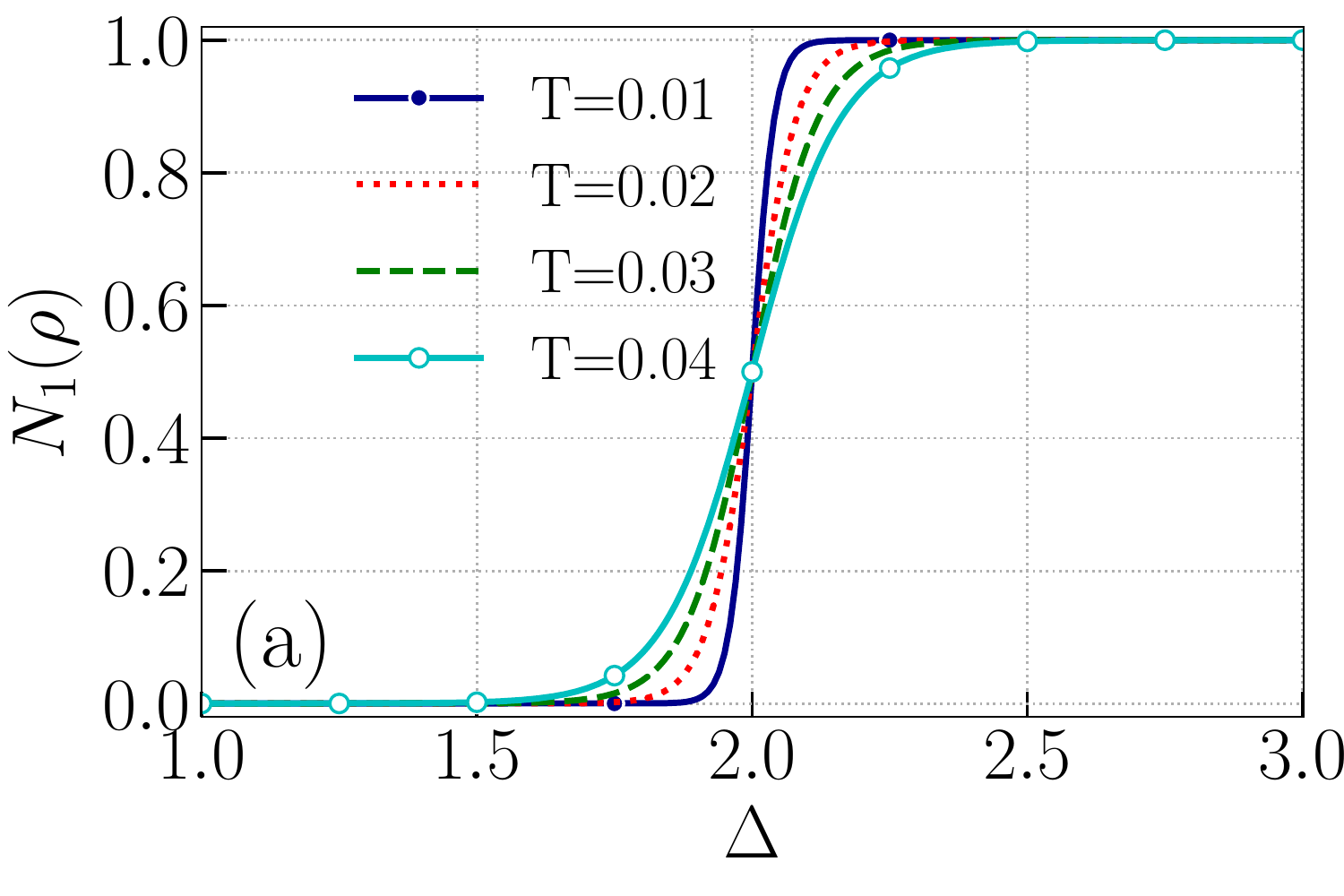}
\centering\includegraphics[width=0.4\linewidth]{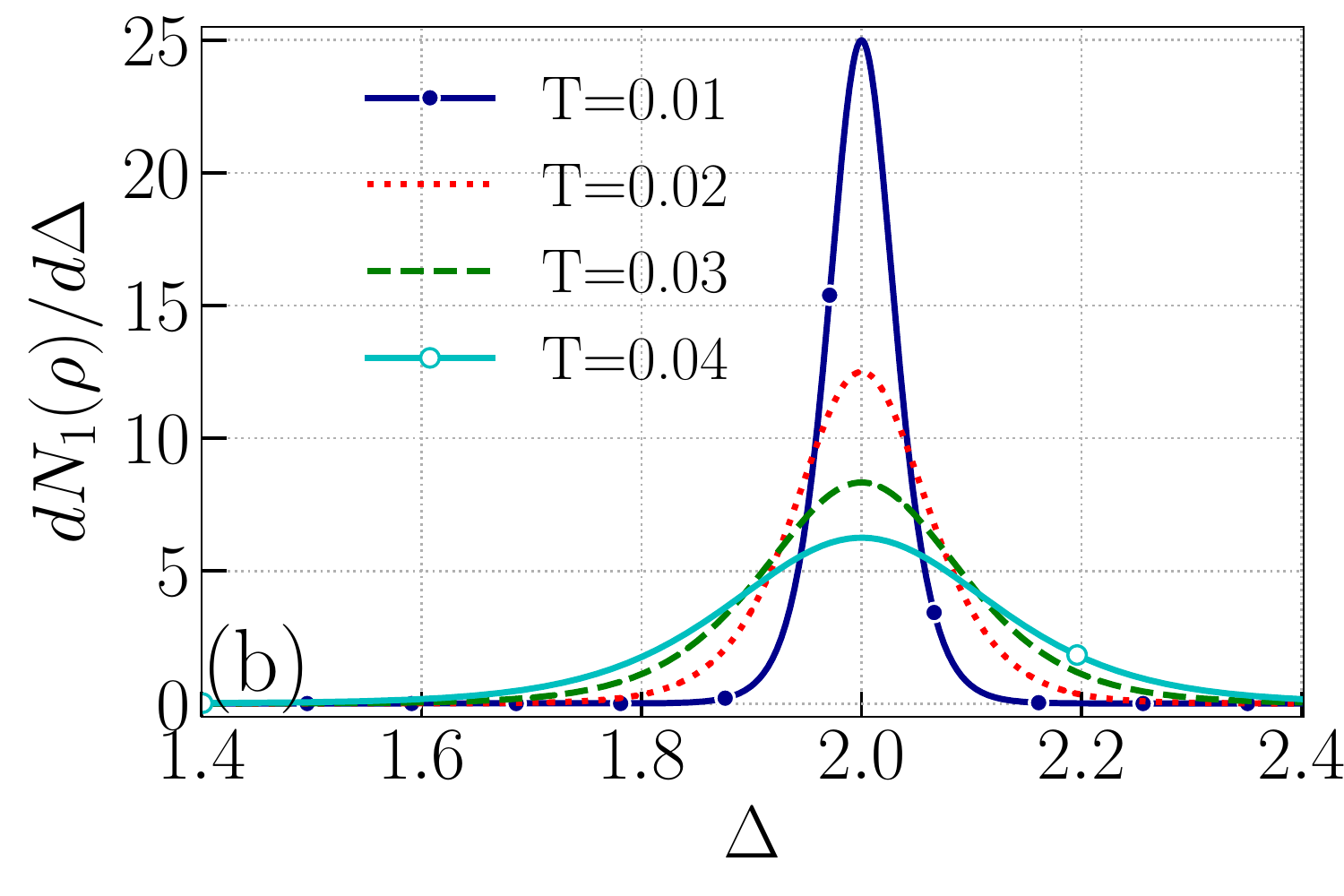}
\caption{(color online) The behaviors of (a) MIN and (b) its derivative as  a function of $\Delta$ with fixed parameters $J=J_1=1$ and $h=2.5$ in the absence of DM interaction for various values of temperatures. For better visibility, we have scaled the magnitude of derivative of MIN with respect to $\Delta$.}
\label{fig1a}
\end{figure*}


\begin{figure*}[!ht]
\centering\includegraphics[width=0.4\linewidth]{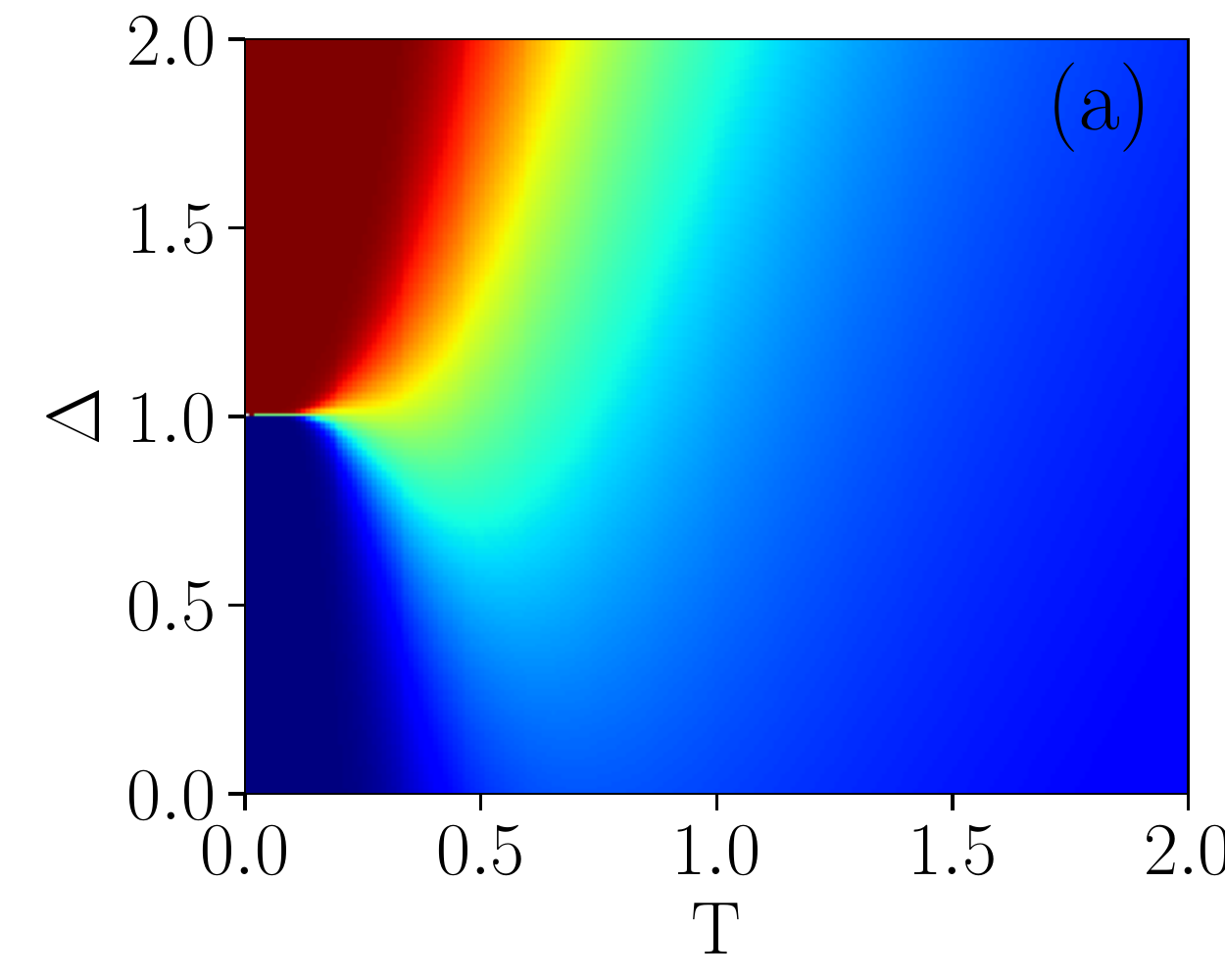}
\centering\includegraphics[width=0.4\linewidth]{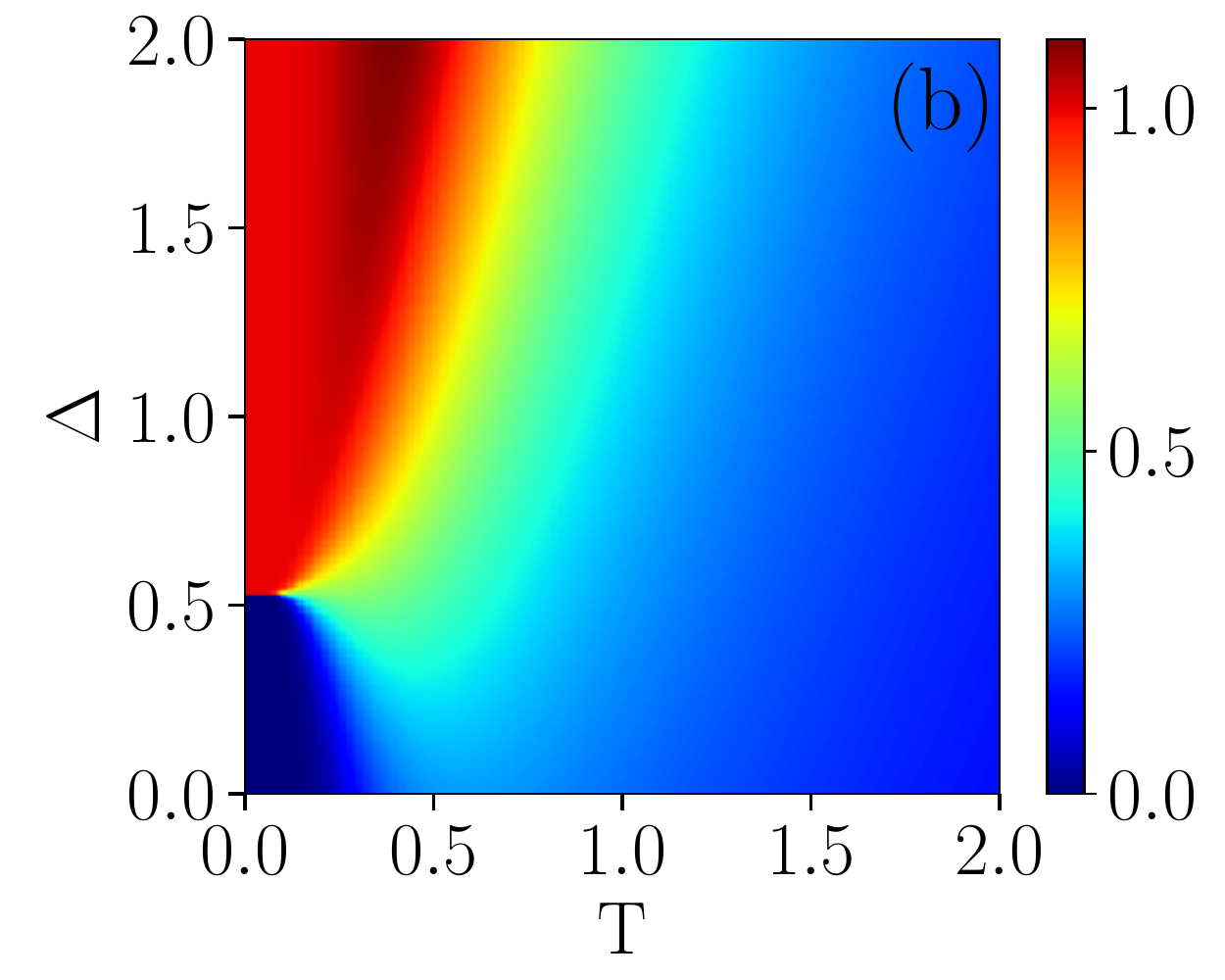}
\caption{(color online) Density of $N_1(\rho)$ as functions of $\Delta$ and temperature for (a) $h=1$, $D=0$ and (b) $h=1$, $D=1$  with the coupling parameters $J = J_1 = 1$.}
\label{density}
\end{figure*}

\begin{figure*}[!ht]
\centering\includegraphics[width=0.4\linewidth]{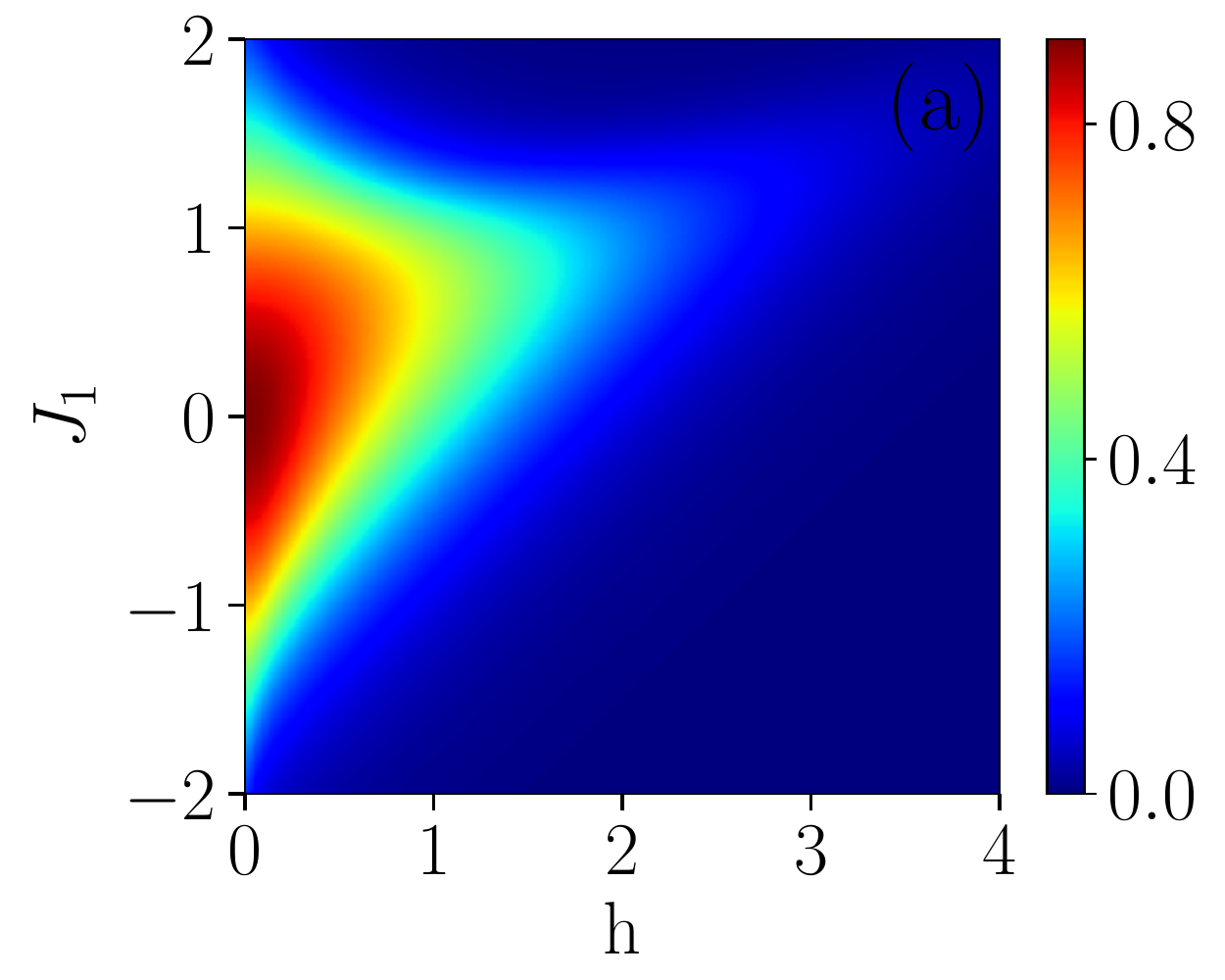}
\centering\includegraphics[width=0.4\linewidth]{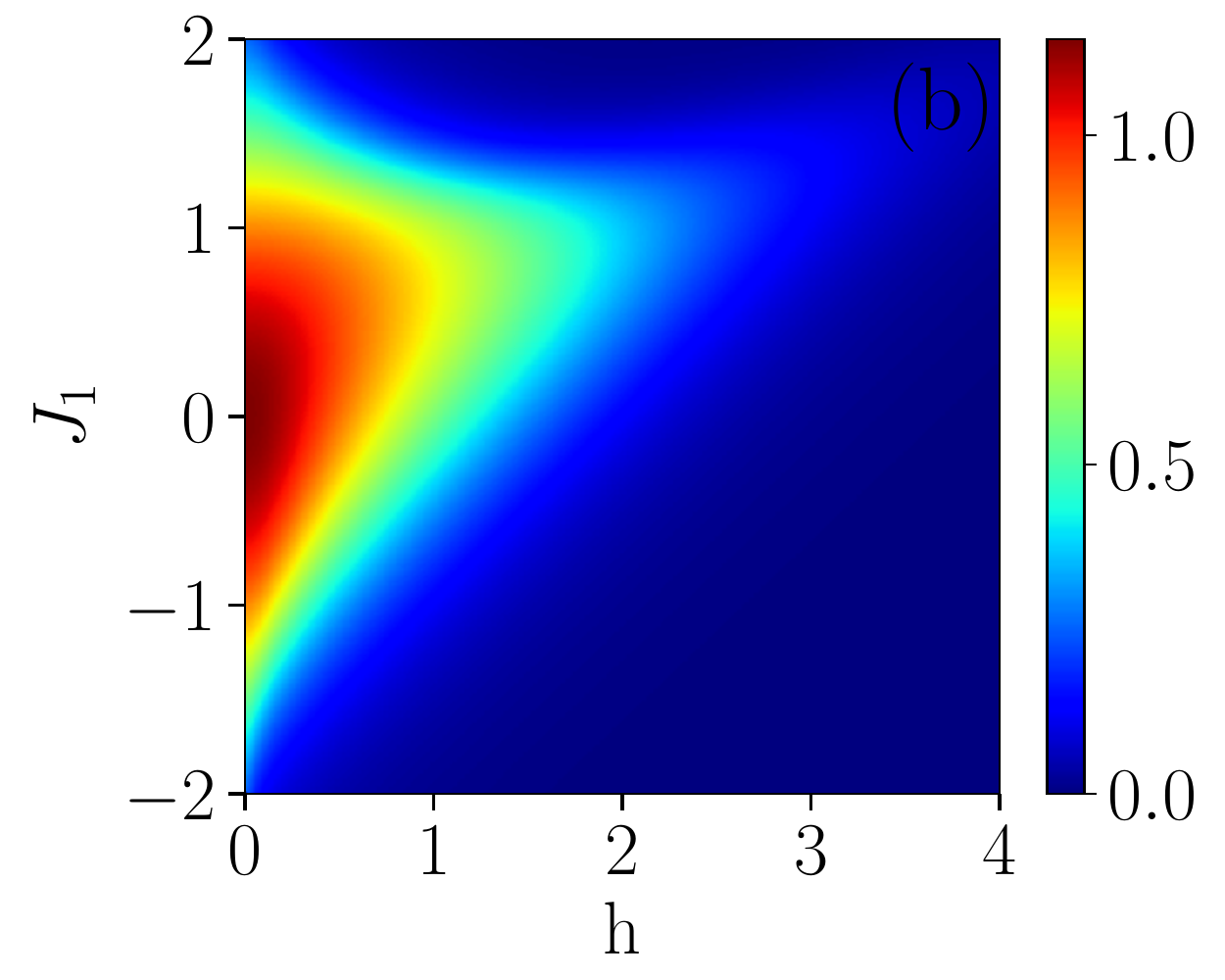}
\caption{(color online) Density of $N_1(\rho)$ as functions of $J_1$ and external magnetic field $h$ for (a) $D=0$ and (b) $D=1$  with the fixed parameters $J=\Delta=1$ and $T=0.5$.}
\label{densityjh}
\end{figure*}
Further, in order to assert that the MIN is a better quantity in capturing quantum correlations, we plot the density of $N_1(\rho)$ as a function of $T$ and $\Delta$ for a fixed value of $h=0$ and $D=0,1$. At $T=0$, $N_1(\rho)=0$ for $\Delta> 1$, it can signal the QPT between the unentangled state in ferrimagnetic phase (UFI) and the entangled state in the quantum ferrimagnetic (ENQ) phase. It is obvious that MIN remains non-zero for the entire parametric space. On the other hand, entanglement captures only lesser quantum correlations  compared to MIN \cite{Rojas2012,Gao2015}, implying that entanglement can grasp only a part of nonlocality in the system. Hence, we conclude that MIN is a better indicator of QPT than entanglement.  Further, from Fig (\ref{fig3}b), we observe that the DM interaction acts as a catalyst to  induce the quantum correlation.

\begin{figure*}[!ht]
\centering\includegraphics[width=0.4\linewidth]{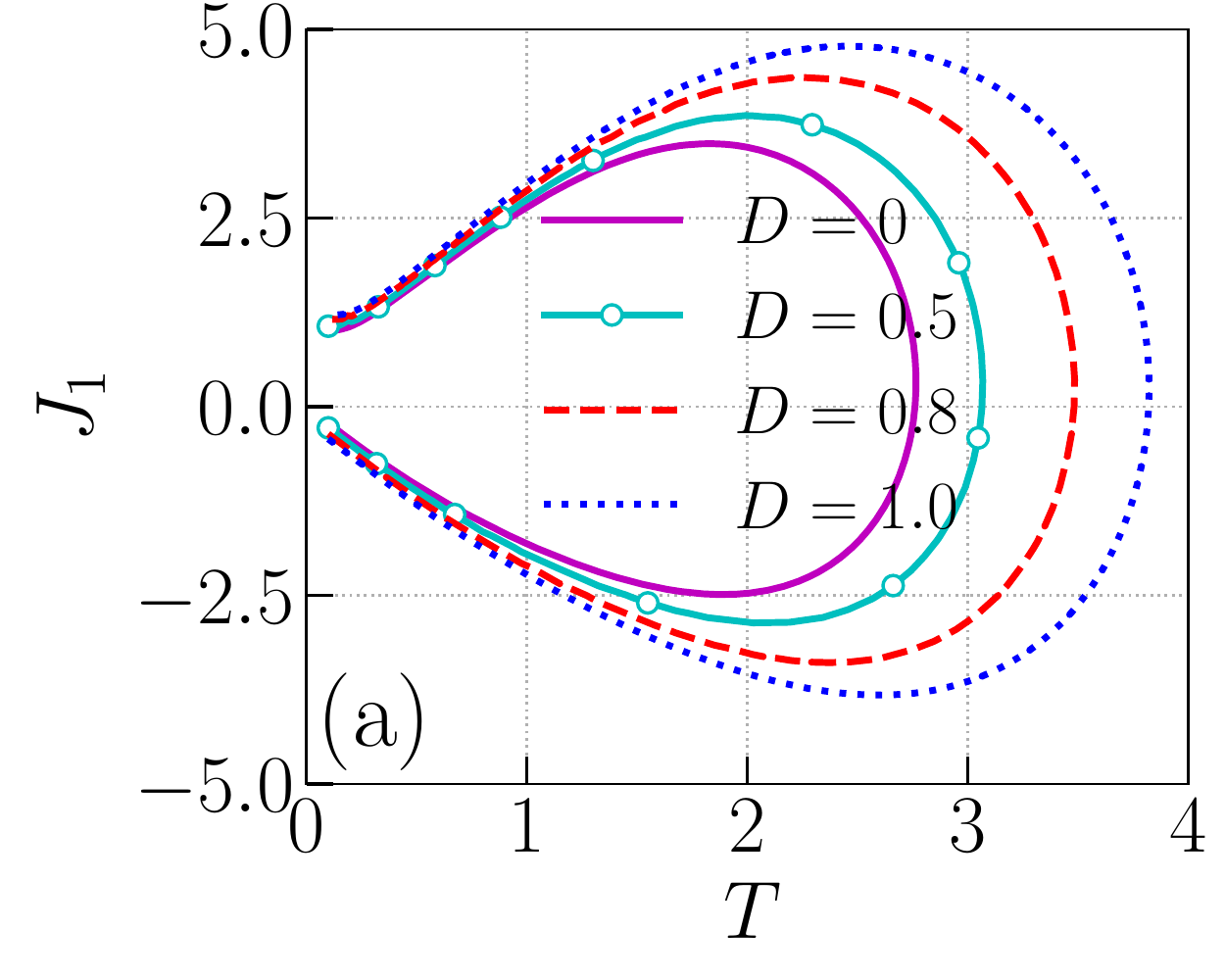}
\centering\includegraphics[width=0.4\linewidth]{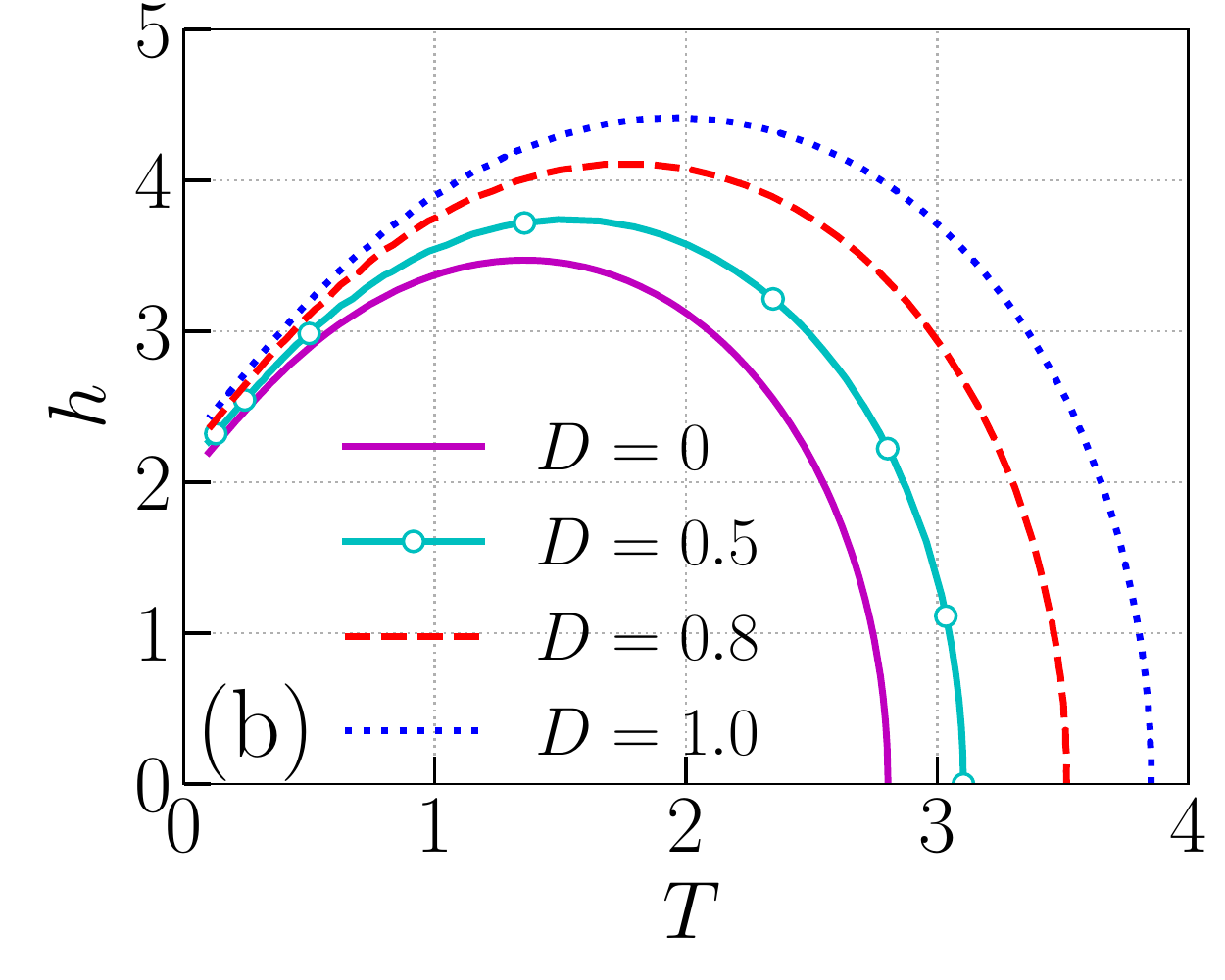}
\caption{(color online) (left) The dependence of the  threshold $J_1$ and threshold temperature of MIN,  (Right) The dependence of the  threshold external magnetic field and threshold temperature of MIN for different DM interaction parameters. Here $\Delta=J=1$.}
\label{PhaseJ1T}
\end{figure*}

 To enhance our understanding of MIN, we depict the density of $N_1(\rho)$ as a function of Ising coupling $J_1$ and the magnetic field $h$ in Fig. (\ref{densityjh}). For $D=0$ (without DM interaction), we observe that the spins are strongly (maximally) correlated for a small region of $J_1$ ($|J_1|\lesssim  1$) in the vicinity of a weak magnetic field. In Rojas et al., it is shown that the region $-1\leq J_1 \leq 1$ corresponds to maximally entangled state \cite{Rojas2012}.  In the present investigation, the MIN is  minimum in the parametric region $|J_1|> 1$ and for higher magnetic field which is identical to that of  Rojas et al., \cite{Rojas2012}. The region of nonzero MIN is limited by the region $|J_1| \lesssim 2$ and $h \lesssim 3$. In addition, to understand the effects of DM interaction, we set $D=1$ and plot the density of MIN as a function of $J_1$ and magnetic field $h$ in Fig. (\ref{densityjh})b. We clearly observe that the introduction of DM interaction enhances MIN and the parametric space corresponding to nonzero MIN remains unaltered.  In Fig. (\ref{PhaseJ1T}), we plot the phase diagram of correlated and uncorrelated  states as a function of Ising coupling parameter $J_1$ and temperature $T$  for  different values of DM interaction parameters ($D=0,0.5,0.8,1$). For $D=0$, in Fig. (\ref{PhaseJ1T}a), it is shown that the purple curve separates the states with zero MIN and nonzero MIN. In other words, the region encompassed by the purple curve corresponds to the state with nonzero MIN (correlated).  In addition, we observe that the area of nonzero MIN region increases with the increase of  DM interaction implying that DM interaction  induces the nonlocal correlation in the system essentially at  higher temperatures.

Figure. (\ref{PhaseJ1T}b) shows the phase diagram of diamond spin chain from the perspective of MIN as a function of magnetic field $h$ and temperature $T$. It indicates the threshold values of magnetic field $h_{\text{th}}$ (the field at which quantum correlations vanish) and temperature $T_{\text{th}}$ (the temperature at which MIN vanish).  Here again, we confirm that the region of nonzero MIN increases with the increase of the DM interaction,  implying  that the enhancement of $D$ can increase the threshold value temperature $T_{\text{th}}$ and thesrhold external magnetic field $h_{\text{th}}$. In other words, the DM interaction creates the correlation between the spins. Further, we observe that the DM interaction induces the correlation in the parametric space where there is no correlation between the spins and strengthens the correlation in the parametric space if the spins are already correlated.
\begin{figure*}[!ht]
\centering\includegraphics[width=0.6\linewidth]{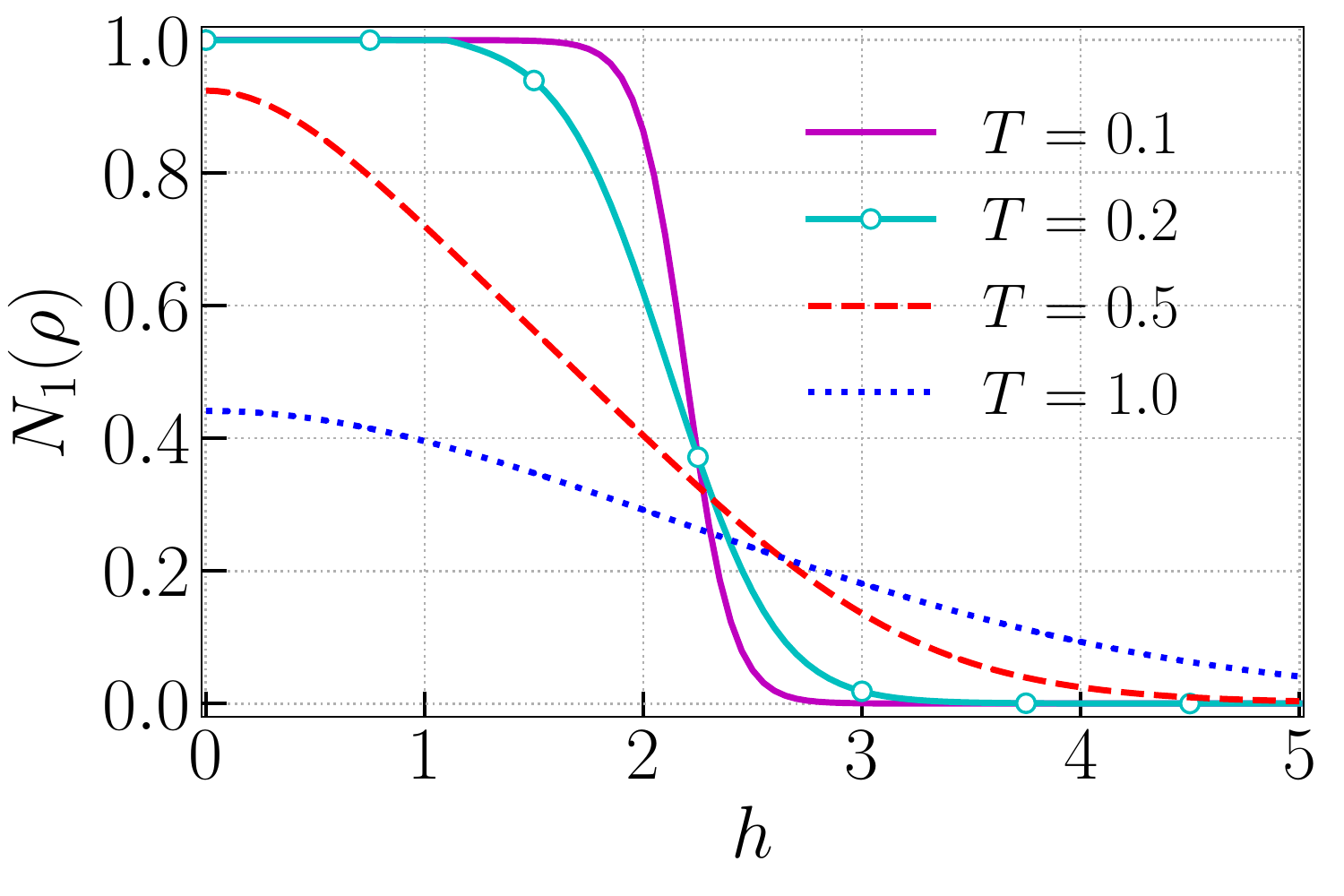}
\caption{(color online) MIN versus the magnetic field $h$ for different temperatures with fixed values $J=J_1=D=\Delta=1$.}
\label{magnetic}
\end{figure*}
Figure (\ref{magnetic}) depicts MIN as a function of applied magnetic field $h$ for different values of temperatures with fixed parameters $\Delta=J=J_1=1$ and $D=1$. For lower temperatures $(T=0.1 ~\text{and}~0.2)$, MIN remains constant in a small range of $h$ and then decreases nonmonotonically with increasing $h$. At higher temperatures $(T=0.5 ~\text{and}~ 1.0)$, the quantum correlation decreases monotonically with external magnetic field and reaches zero gradually. The application of magnetic field decreases the MIN. Here also, we observe that the increase of temperature tends to reduce the quantum correlation in the system.

\section{Conclusion}
To summarize, we have studied the behaviors of thermal quantum correlations captured by measurement-induced nonlocality (MIN) in a spin-1/2 Ising-Heisenberg diamond infinite spin chain. It is shown that MIN has an advantage in spotlighting the quantum critical phenomena in the system. At low temperatures, depending on the external magnetic field, MIN identifies the phase transition from an unentangled state in the ferrimagnetic phase to an entangled state in the ferrimagnetic  phase $(h\geq 2)$, from an entangled state in the ferrimagnetic phase to an unentangled ferromagnetic state (with $h<2 $). For high temperatures, the quantum correlation vanishes and does not exhibit critical phenomena, implying that thermal fluctuation dominates the quantum effects in the system.  The DM interaction essentially induces the nonlocality in the system and reduces the role of anisotropy parameter in initiating phase transition. The intervention of the magnetic field decreases the nonlocality of the system and DM  interaction greatly enhances the quantum correlation between the spins. 
\noindent

\section*{Acknowledgment}
 Authors thank the reviewer's for their critical comments to improve the contents of the paper. SB and RR thank the Council of Scientific and Industrial Research (CSIR), Government of India for the financial support under Grant No. 03(1456)/19/EMR-II.  RM acknowledge  the financial support from the Council of Scientific and Industrial Research (CSIR), Government of India, under Grant No. 03(1444)/18/EMR-II. RR wishes to acknowledge the DAE-NBHM for the financial support under the scheme 02011/3/20/2020-R\&D-II.


\section*{Appendix}

 An arbitrary state in the Bloch representation can be written as 
\begin{equation}
\rho=\frac{1}{2}\left[ X_0 \otimes Y_0+\sum_{i=1}^3 x_i (X_i\otimes Y_0)+\sum_{j=1}^3 y_j (X_0 \otimes Y_j)+\sum_{i,j\neq 1} t_{ij} X_i \otimes Y_j\right] \label{Equations} 
\end{equation}
where $x_i=\text{Tr}(\rho(X_i\otimes Y_0))$, $y_j=\text{Tr}(\rho(X_0 \otimes Y_j))$ are the components of Bloch vector and $t_{ij}=\text{Tr}(\rho(X_i \otimes Y_j))$ being real matrix elements of correlation matrix $T$. Orthonormal operators in respective state spaces are $\{X_0,X_1,X_2,X_3 \}=\{\mathds{1},\sigma_1,\sigma_2,\sigma_3 \}/\sqrt{2} $ and $\{Y_0,Y_1,Y_2,Y_3 \}=\{\mathds{1},\sigma_1,\sigma_2,\sigma_3 \}/\sqrt{2} $, where $\sigma_i$ are the Pauli matrices. Then, the closed formula of Hilbert-Schmidt norm based MIN (\ref{MIN2}) is given as  \cite{Luo2011}
\begin{equation}
N_2(\rho ) =
\begin{cases}
\text{Tr}(TT^t)-\frac{1}{\| \textbf{x}\| ^2}\textbf{x}^tTT^t\textbf{x}& 
 \text{if} \quad \textbf{x}\neq 0,\\
 \text{Tr}(TT^t)- \tau_{\text{min}}&  \text{if} \quad \textbf{x}=0,
\end{cases}
\label{HSMIN}
\end{equation}
where $\tau_{\text{min}}$ is the least eigenvalue of matrix $TT^t$, the superscripts $t$ stands for the transpose and the vector $\textbf{x}=(x_1,x_2,x_3)^t$.

 Without loss of generality, we can rewrite  Eq. (\ref{Equations}) as 
\begin{equation}
\rho=\frac{1}{4}\left[ \mathds{1} \otimes \mathds{1}+\sum_{i=1}^3 x_i (\sigma_i\otimes \mathds{1})+\sum_{j=1}^3 y_j (\mathds{1} \otimes \sigma_j)+\sum_{i}c_{i} \sigma_i \otimes \sigma_i\right] \label{EQ} 
\end{equation}
where $c_{i}=\text{Tr}(\rho(\sigma_i\otimes \sigma_i))$.  The closed formula of $N_1(\rho)$ is given as 
\begin{equation}
N_1(\rho)=
\begin{cases}
\frac{\sqrt{\chi_+}~+~\sqrt{\chi_-}}{2 \Vert \textbf{x} \Vert_1} & 
 \text{if} \quad \textbf{x}\neq 0,\\
\text{max} \lbrace \vert c_1\vert,\vert c_2\vert,\vert c_3\vert\rbrace &  \text{if} \quad \textbf{x}=0,
\end{cases}
\label{TDMIN}
\end{equation}
where $\chi_\pm~=~ \alpha \pm 2 \sqrt{\tilde{\beta}} \Vert \textbf{x} \Vert_1 ,\alpha =\Vert \textbf{c} \Vert^2_1 ~\Vert \textbf{x} \Vert^2_1-\sum_i c^2_i x^2_i,\tilde{\beta}=\sum_{\langle ijk \rangle} x^2_ic^2_jc^2_k, \vert c_i \vert $ is the absoulte value of $c_i$ and the summation runs over cyclic permutation of $\lbrace 1,2,3 \rbrace$.
Using the following matrices, 
\begin{align}
T = 
\begin{pmatrix}
 2\rho_{23} & 0 & 0  \\
 0 &  2\rho_{23} & 0  \\
 0 &  0 & \rho_{11}-\rho_{22}-\rho_{33}+\rho_{44}
\end{pmatrix},  \nonumber~~~~~ 
\text{x} = 
\begin{pmatrix}
0 &  0 & \rho_{11}+\rho_{22}-\rho_{33}-\rho_{44}
\end{pmatrix}^t,~~~~~~~~~~~~~~~~~~~~~~~~~~~ \nonumber
\end{align}

and
\begin{align}
\text{c} = 
\begin{pmatrix}
2\rho_{23} & 2\rho_{23} & \rho_{11}-\rho_{22}-\rho_{33}+\rho_{44}
\end{pmatrix}^t
\label{matrices},
\end{align}
the MIN and trace norm MIN are calculated from Eqs. (\ref{HSMIN}) and (\ref{TDMIN}) as
\begin{align}
N_2(\rho)=2~ \lvert \mathcal{\rho}_{2,3}\rvert^2 ~~ \text{and} ~~ N_1(\rho)=2~ \lvert \mathcal{\rho}_{2,3}\rvert.
\end{align}

\end{document}